\documentclass[aps,physrev,twocolumn,superscriptaddress,amsmath,amssymb]{revtex4-2}

\usepackage{graphicx}
\usepackage{amsmath,amssymb}
\usepackage{bm}
\usepackage{hyperref}
\usepackage{braket}
\usepackage{tikz}
\usepackage{quantikz}
\usepackage{xcolor}
\usepackage[normalem]{ulem}
\newcommand{\Hcnot}{H_{\scriptstyle\mathrm{CNOT}}}

\newcommand{\op}[1]{\operatorname{#1}}

\begin{document}

\title{Topological sum rule for geometric phases of quantum gates}

\author{Nadav Orion}
\affiliation{Department of Physics, Technion, Israel Institute of Technology, Haifa 3200003, Israel}

\author{Boris Rotstein}
\affiliation{Department of Physics, Technion, Israel Institute of Technology, Haifa 3200003, Israel}

\author{Nirron Miller}
\affiliation{Department of Physics, The University of Texas, Austin, Texas 78712}

\author{Eric Akkermans}
\email{eric@physics.technion.ac.il}
\affiliation{Department of Physics, Technion, Israel Institute of Technology, Haifa 3200003, Israel}

\date{\today}

\begin{abstract}
We establish a topological sum rule, $\nu_U = \frac{1}{2\pi}\sum_n\gamma_n = m\nu_H$, connecting the geometric phases accumulated by a two-qubit system over a complete basis of initial states to the winding number $\nu_H$ classifying its Hamiltonian. Implementations of the same gate from different topological classes must distribute these phases differently, making their distinction measurable through the Wootters concurrence. As a corollary, nontrivial topology is a necessary condition for entanglement: only Hamiltonians with access to $\nu_H \neq 0$ can generate it.
\end{abstract}

\maketitle

\section{Introduction}
\label{sec:intro}

Quantum entanglement and topological phases in condensed matter have become increasingly interconnected. We build on the recent discovery that quantum entanglement and the topology of a Hilbert space are fundamentally intertwined and not just conveniently related. Specifically, two qubits without topological features cannot be entangled~\cite{Orion2025}. This paper offers an interdisciplinary perspective and methodology for studying two-qubit gates, based on an approach rooted in topological condensed matter.

The same quantum gate can be implemented by physically distinct Hamiltonians. Entangling two-qubit Hamiltonians fall into topologically disconnected classes, labeled by a Hamiltonian winding number $\nu_H$ \cite{Orion2025}. We show that geometric phases generated by implementations from different classes must redistribute across initial states in a way constrained by the topological sum rule
\begin{equation}
\nu_U \;=\; \frac{1}{2\pi}\sum_n\gamma_n \;=\; m\,\nu_H\,.
\label{eq:sumrule_intro}
\end{equation}
Here $\nu_U$ is the $\mathbb{Z}$-valued Chern invariant of the evolution operator $U(t)$, i.e.\ the winding number of $\det U(t)$ around the origin in $\mathbb{C}$, $\{\gamma_n\}$ are the geometric phases over a complete set of initial states, and $m$ is the number of gate cycles. This redistribution is directly measurable, and its dependence on entanglement follows from the interplay between the Wootters concurrence and the two-qubit geometric phase. As a corollary, nonvanishing $\nu_H$ is a necessary condition for entanglement.

Entangling two-qubit Hamiltonians, classified by Cartan decomposition~\cite{Orion2025,Altland1997,Kitaev2009}, take the form
% Cartan coupling symbols (c_x,c_y,c_z) distinct from Schmidt state parameters (\alpha,\beta)
% and geometric phase \gamma used below.
\begin{equation}
H_e(c_x,c_y,c_z) = c_x\,\sigma_x\otimes\sigma_x + c_y\,\sigma_y\otimes\sigma_y + c_z\,\sigma_z\otimes\sigma_z\,.
\label{eq:He}
\end{equation}
The Hamiltonian topological number~\cite{Orion2025}
\begin{equation}
\nu_H = \frac{1}{2}(n_+ - n_-)
\label{eq:nuH}
\end{equation}
counts the signed imbalance of positive ($n_+$) and negative ($n_-$) eigenvalues, partitioning the space of entangling Hamiltonians into three disconnected sectors (Fig.~\ref{fig:topology}a). Distinct choices of $(c_x,c_y,c_z)$ can produce the same overall unitary $U(H_e)=e^{-iH_e T}$ up to local transformations, yet belong to different topological sectors; their geometric phases must then differ.

\section{Topological Sum Rule}
\label{sec:sumrule}

We now derive~\eqref{eq:sumrule_intro}. Note that this derivation is applicable for any number of qubits. Consider the quantity~\cite{Baker1958,Bolle1987,Akkermans1997,Akkermans2000}
\begin{equation}
\nu_U = \frac{i}{2\pi}\left[\int_0^T\frac{d}{dt}\ln(\det U)\,dt - \ln\det\!\left(U^\dagger(0)\,U(T)\right)\right]
\label{eq:nuU}
\end{equation}
where $T$ is the period of the evolution, whose determinant contour $z(t)=\det U(t)$ lies on the unit circle. By the Cauchy integral formula its winding number around $z=0$ is an integer:
\begin{equation}
\nu_U = \frac{i}{2\pi}\,n\oint\frac{dz}{z} = -n\,.
\label{eq:nuU_proof}
\end{equation}
This proof and result holds regardless of periodicity in $U$ (see Appendix). Using $\frac{d}{dt}\ln(\det U)=\mathrm{Tr}\!\left(U^\dagger\frac{d}{dt}U\right)$ and expanding the trace over any complete basis $\{|\psi_n\rangle\}$, this expression becomes precisely the sum rule~\eqref{eq:sumrule_intro}, i.e.\
\begin{equation}
\nu_U =\frac{1}{2\pi}\sum_n\gamma_n\,,
\label{eq:nuU_gamma}
\end{equation}
where $\gamma_n$ is the geometric phase~\cite{Berry1984,Simon1983,Aharonov1987} of the state $|\psi_n\rangle$ over the period $T$. See Appendix for a discussion of gauge invariance and calculation examples. The geometric phase of a state $\ket{\psi(t)}$ is given by
\begin{equation}
\gamma = i\oint_\Gamma \langle\psi|d\psi\rangle + \arg\langle\psi(0)|\psi(T)\rangle\,.
\label{eq:gamma}
\end{equation}
$\gamma$ as obtained from the Berry connection $\mathcal{A}=\langle\psi|d\psi\rangle$~\footnote{$\mathcal{A}$ is the connection 1-form of the $U(1)$ fibration.}, is gauge-invariant and determined solely by the path $\Gamma$ in Hilbert space \cite{Mukunda_Simon_1993,Vanderbilt_book_2018,Cohen2019,Cisowski2022}.

While each individual $\gamma_n$ depends on the initial state and the specific Hamiltonian path, their sum is a topological invariant fixed by the class of $H$. While analogous sum rules govern Zak phases in band theory~\cite{Zak1989}, the present result applies to the evolution operator of a controllable quantum gate and directly connects the observable geometric phase to the implementation-independent topological class of the gate Hamiltonian, a connection with no counterpart in the condensed matter setting.

\begin{figure}[t]
\centering
\includegraphics[width=0.8\columnwidth]{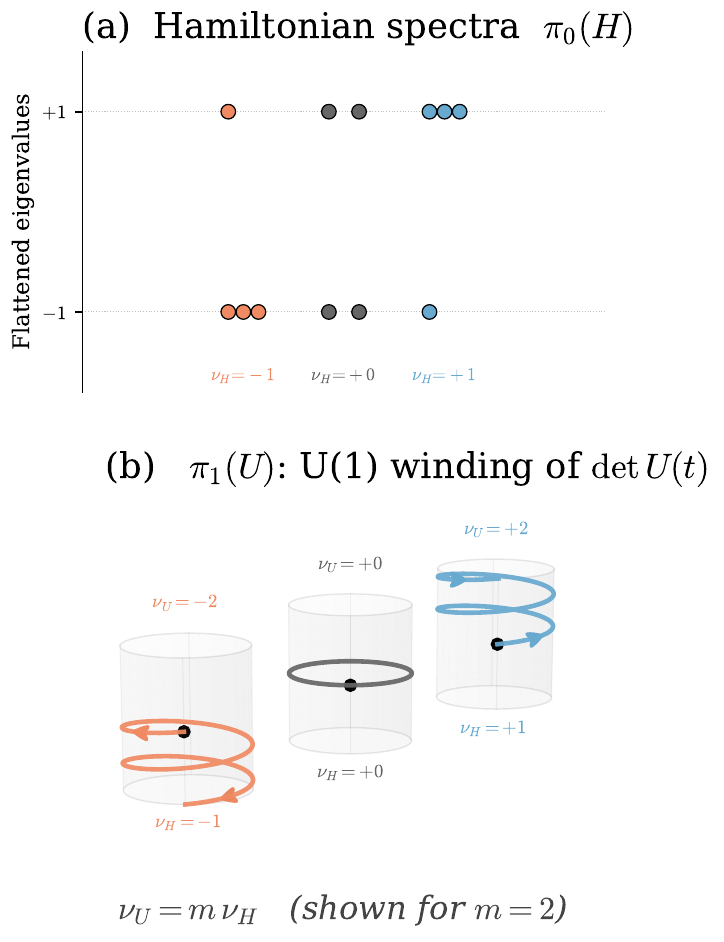}
\caption{Topological correspondence $\pi_0(H) = \pi_1(U)$. (a)~Flattened eigenvalue spectra of two-qubit Hamiltonians: each column shows how the four eigenvalues ($\pm 1$) distribute, with the imbalance $\nu_H = (n_+ - n_-)/2$ labeling the three disconnected classes for Hamiltonians of the type~\eqref{eq:He}. (b)~Phase evolution of $\det U(t) = e^{-2i\nu_H t}$ as a spiral, projected to a circle when taken modulo $2\pi$. Each $\nu_H$ produces $\nu_U = m\nu_H$ windings (shown for $m=2$), making the correspondence between Hamiltonian classes and loop topology visually manifest.}
\label{fig:topology}
\end{figure}

Evaluating~\eqref{eq:nuU} on the flattened representative of each class (eigenvalues $\pm 1$) gives $\det U(t) = e^{-2i\nu_H t}$, which winds $m\nu_H$ times over $m$ cycles, establishing
\begin{equation}
\nu_U = m\,\nu_H\,,
\label{eq:nuU_nuH}
\end{equation}
the topological correspondence $\pi_0(H)=\pi_1(U)$ illustrated in Fig.~\ref{fig:topology}. Although $\nu_U$ depends on the cycle count $m$, the Hamiltonian class $\nu_H=\nu_U/m$ is uniquely recovered; $\nu_U$ is thus the directly measurable geometric-phase signature of $\nu_H$, related to it by a known factor set by the experimental protocol. For $\nu_H\neq 0$, $\nu_U$ ranges over all integer multiples of $\nu_H$, corresponding to the $\mathbb{Z}$ invariant~\cite{Eguchi1980}. See also Appendix.

The sum rule~\eqref{eq:nuU_gamma} has an immediate consequence for gate implementation: two implementations of the same gate whose Hamiltonians belong to \emph{different} topological classes must produce geometric phases that redistribute across initial states while keeping their sum fixed at $2\pi\nu_U$. This redistribution is observable through the two-qubit geometric phase, which depends on the Wootters concurrence $C_{2q}\in[0,1]$~\cite{Hill1997,Wootters1998}, the standard measure of entanglement~\cite{Haroche2006,Horodecki2009,Paneru2020}:
\begin{equation}
C_{2q}(t) \equiv |\langle\psi(t)|\,\Theta\,|\psi(t)\rangle|\,,
\label{eq:concurrence}
\end{equation}
where $\Theta = (\sigma_y\otimes\sigma_y)K$ is the time-reversal operator, $K$ is complex conjugation in the computational basis $|jk\rangle$ ($j,k\in\{0,1\}$) with $K|jk\rangle=|jk\rangle$ and $KiK=-i$~\cite{Bell1964,Clauser1969,Verstraete2002}. Concurrence has been measured in multiple platforms~\cite{Romero2007,Walborn2006,Satoor2021}.

\section{Two-Qubit Geometric Phase and Concurrence}
\label{sec:gamma2q}

A general two-qubit state (see Supplementary Material \cite{SM}) takes the Schmidt-sphere form~\cite{Sjoqvist2000,Bertlmann2004}
\begin{equation}
|\psi_{2q}\rangle = \cos\frac{\alpha}{2}\,e^{-i\beta/2}|\hat{n}_1,\hat{n}_2\rangle + \sin\frac{\alpha}{2}\,e^{i\beta/2}|-\hat{n}_1,-\hat{n}_2\rangle
\label{eq:psi2q}
\end{equation}
where the states $|\pm\hat{n}_i\rangle$ denote the usual spin-$\frac{1}{2}$ basis along $\hat{n}_i\left(\theta_i,\phi_i\right)$ on the Bloch sphere (i.e.\ eigenstates of $\hat{n}_i\cdot\vec{\sigma}$), and $C_{2q} = \sin\alpha$. The Schmidt decomposition establishes the ranges $\alpha\in[0,\pi/2]$, $\theta_{1,2}\in[0,\pi]$, and $(\beta,\phi_{1,2})\in[0,2\pi]$. The angle $\alpha$ is thus uniquely (one-to-one) related to the concurrence. 

Although the entanglement dependence of the two-qubit geometric phase was established in~\cite{Sjoqvist2000}, the Schmidt-sphere form makes the topological classification via $\nu_H$ transparent. Applying~\eqref{eq:gamma} to~\eqref{eq:psi2q} yields (see Supplementary Material \cite{SM}):
\begin{equation}
\gamma_{2q} = \frac{1}{2}\int_0^T dt\,\sqrt{1-C_{2q}^2}\left(\dot{\beta}+\dot{\phi}_1\cos\theta_1 + \dot{\phi}_2\cos\theta_2\right).
\label{eq:gamma2q}
\end{equation}
The single-qubit phase $\int dt\dot{\phi}\cos\theta$~\cite{Aharonov1987} appears once per qubit, each suppressed by $\sqrt{1-C_{2q}^2}=\cos\alpha$, consistent with $\langle\vec{\sigma}^{(i)}\rangle = \hat{n}_i\cos\alpha$~\eqref{eq:psi2q}. The extra non-local term $\sqrt{1-C_{2q}^2}\,\dot{\beta}$, with no single-qubit analog, encodes interactions. When $C_{2q}=0$, \eqref{eq:gamma2q} reduces to two independent single-qubit phases~\footnote{The $\dot{\beta}$ term in~\eqref{eq:gamma2q} vanishes, as in that case~\eqref{eq:psi2q} is a separable state, and $\beta$ is a global phase.}. See Supplementary Material~\cite{SM} for a derivation and discussion of both~\eqref{eq:psi2q} and~\eqref{eq:gamma2q} based on the method introduced in~\cite{Bouchiat1988}.

\section{Gate Discrimination}
\label{sec:discrimination}

\subsection{SWAP gate}
\label{sec:swap}

Consider two physical realizations of the SWAP gate,
\begin{equation}
\mathrm{SWAP}|jk\rangle = |kj\rangle, \qquad \mathrm{SWAP}^2 = I\,.
\label{eq:SWAP}
\end{equation}
The first uses the Heisenberg Hamiltonian~\cite{Loss1998,Burkard2023,Krantz2019}
\begin{equation}
H_1 = \lambda\,\vec{\sigma}^{(1)}\cdot\vec{\sigma}^{(2)} = \lambda(\sigma_x\otimes\sigma_x + \sigma_y\otimes\sigma_y + \sigma_z\otimes\sigma_z)\,,
\label{eq:H2}
\end{equation}
\begin{equation}
\mathrm{SWAP}_1 = e^{i\pi/4}\exp\!\left(-iH_2\frac{\pi}{4\lambda}\right).
\label{eq:SWAP1}
\end{equation}

\begin{figure}[t]
\centering
$\mathrm{SWAP}_2 =\;$
\begin{quantikz}
\qw & \ctrl{1} & \targ{} & \ctrl{1} & \qw \\
\qw & \targ{}  & \ctrl{-1} & \targ{} & \qw
\end{quantikz}
\caption{A quantum circuit implementing SWAP using three CNOT operators, where for the first and last operators qubit A controls qubit B, and for the second qubit B controls qubit A (reversed).}
\label{fig:swap_circuit}
\end{figure}
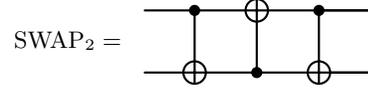

The second uses three sequential CNOT gates (Fig.~\ref{fig:swap_circuit}):
\begin{equation}
    \op{CNOT}\ket{jk} = |k\rangle\otimes\ket{j\oplus k}, \qquad \mathrm{CNOT}^2 = I
\end{equation}
with $\oplus$ addition modulo 2.
CNOT is implemented in this paper via the cross-resonance Hamiltonian~\cite{Krantz2019,Rigetti2010,Malekakhlagh2020,Magesan2020}:
\begin{equation}
\Hcnot = \frac{w}{2}(\sigma_z\otimes I + I\otimes\sigma_x - \sigma_z\otimes\sigma_x)\,,
\label{eq:HCNOT}
\end{equation}
\begin{align}
\mathrm{SWAP}_2 &= \mathrm{CNOT}\cdot\mathrm{CNOT}_\leftrightarrow\cdot\mathrm{CNOT}\,,\nonumber\\
\mathrm{CNOT} &= \exp\!\left(-i\Hcnot\frac{\pi}{2w}\right).
\label{eq:SWAP2}
\end{align}
The resulting geometric phase is path-dependent and additive across the gate segments (see Supplementary Material \cite{SM} for proof that abrupt vs.\ ramped switching gives identical results).

\begin{figure}[t]
\centering
\includegraphics[width=\columnwidth]{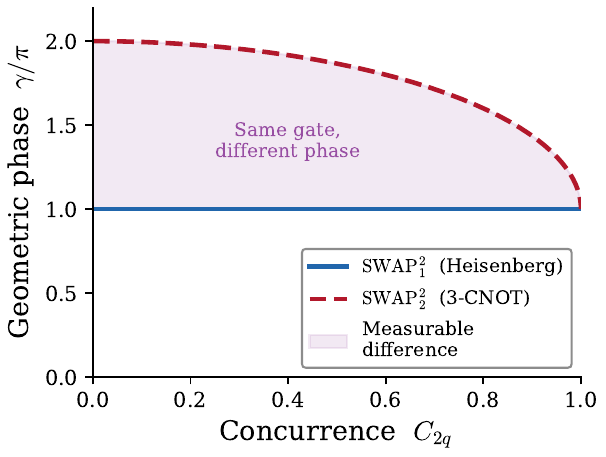}
\caption{Geometric phase $\gamma/\pi$ as a function of concurrence $C_{2q}$ for two implementations of the SWAP$^2$ gate applied to symmetric initial states~\eqref{eq:psi_s}. The Heisenberg-based SWAP$_1$ (solid blue) yields a constant phase $\gamma=\pi$, independent of entanglement. The 3-CNOT-based SWAP$_2$ (dashed red) decreases from $2\pi$ at $C_{2q}=0$ to $\pi$ at maximal entanglement. The shaded region highlights the measurable difference between two implementations of the same gate.}
\label{fig:swap_phases}
\end{figure}

\begin{figure}[t]
\centering
\includegraphics[width=0.85\columnwidth]{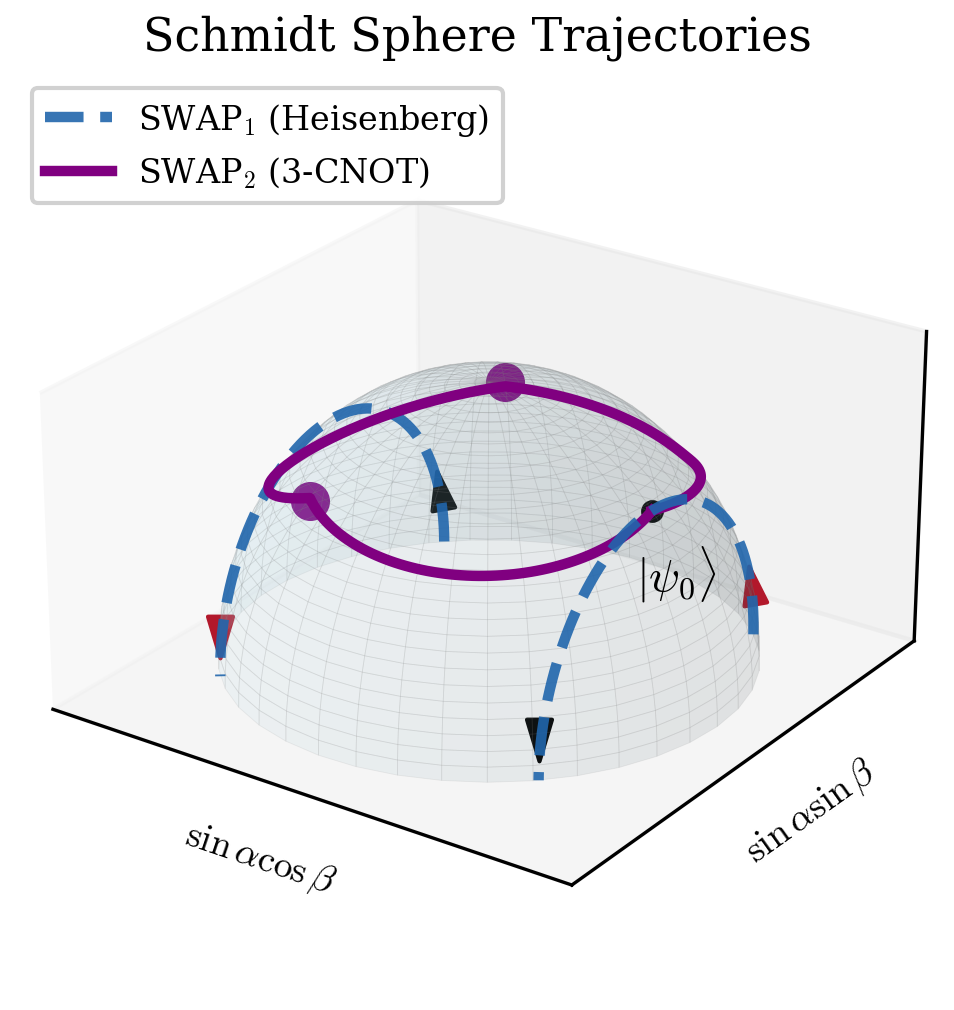}
\caption{Two cyclic paths on the Schmidt sphere, both starting and ending at the same initial state $|\psi_0\rangle$, corresponding to two distinct implementations of the SWAP gate. The Heisenberg-based SWAP$_1$ (dashed blue) traces a smooth great-circle arc. The 3-CNOT SWAP$_2$ (solid purple) follows three distinct segments, with circles marking the corners where the Hamiltonian switches; these segments correspond to the three CNOT operations in~\eqref{eq:SWAP2}. The arrows on the SWAP$_1$ curve show how the path continues from one side of the sphere to the other. Although both paths implement the same gate, they enclose different solid angles and thus acquire different geometric phases; this is a direct manifestation of their different topological classes.}
\label{fig:schmidt_paths}
\end{figure}

The two Hamiltonians belong to distinct topological classes: the $\mathrm{SWAP}_1$ implementation is time-reversal symmetric, whereas $\mathrm{SWAP}_2$ breaks time-reversal symmetry. Applied to the symmetric initial state
\begin{equation}
|\psi_s(t\!=\!0)\rangle = \cos\frac{\alpha_0}{2}\,e^{-i\beta_0/2}|00\rangle + \sin\frac{\alpha_0}{2}\,e^{i\beta_0/2}|11\rangle
\label{eq:psi_s}
\end{equation}
over two consecutive SWAP gates (a full cycle), the sum rule demands their per-state phases differ. Explicitly,
\begin{equation}
\gamma(\mathrm{SWAP}_1^2) = \pi\,,\qquad \gamma(\mathrm{SWAP}_2^2) = \pi(1+\cos\alpha_0)\,.
\label{eq:SWAP_phases}
\end{equation}
SWAP$_1$ gives a constant phase regardless of entanglement; SWAP$_2$ gives a concurrence-dependent phase (Fig.~\ref{fig:swap_phases}). On antisymmetric states~\eqref{eq:psi_as}
\begin{equation}
|\psi_{\mathrm{as}}(t\!=\!0)\rangle = \cos\frac{\alpha_0}{2}\,e^{-i\beta_0/2}|10\rangle + \sin\frac{\alpha_0}{2}\,e^{i\beta_0/2}|01\rangle\,.
\label{eq:psi_as}
\end{equation}
both phases coincide:
\begin{equation}
\gamma_{\op{SWAP}^2}^{(\mathrm{as})} = 2\pi\sin\alpha_0\cos\beta_0 = 2\pi\,C_{2q}(t\!=\!0)\cos\beta_0,
\label{eq:gamma_H2}
\end{equation}
showing that for both implementations $\nu_U=1$. While fitting the result $\nu_{H_1}=1$, $\nu_U=1$ is not immediately expected for $\mathrm{SWAP}_2^2$ as $\nu_{\Hcnot}=-1$. The change in topological number is a result of the transformation between $\Hcnot$ and $H_{\scriptstyle\mathrm{CNOT}_\leftrightarrow}$ mixing between states with positive and negative eigenvalues, therefore changing the topology during the piecewise time evolution. Fig.~\ref{fig:schmidt_paths} illustrates the geometric origin: the two implementations trace topologically distinct paths on the Schmidt sphere, enclosing different solid angles.

\subsection{CNOT gate}
\label{sec:cnot}

The topological redistribution for CNOT is illustrated by comparing implementation via~\eqref{eq:HCNOT} with
\begin{equation}
\mathrm{CNOT}_2 = (\op{H}\otimes \op{H})\,\mathrm{CNOT}_\leftrightarrow\,(\op{H}\otimes \op{H})\,,
\label{eq:CNOT2}
\end{equation}

\begin{figure}[b]
\centering
$\mathrm{CNOT}_2 =\;$
\begin{quantikz}
\qw & \gate{\op{H}} & \targ{}   & \gate{\op{H}} & \qw \\
\qw & \gate{\op{H}} & \ctrl{-1} & \gate{\op{H}} & \qw
\end{quantikz}
\caption{A quantum circuit implementing CNOT (qubit A controls qubit B) using Hadamard operators and CNOT$_\leftrightarrow$ (qubit B controls qubit A).}
\label{fig:cnot_circuit}
\end{figure}
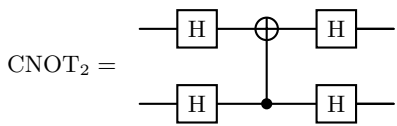
\begin{figure}[t]
\centering
\includegraphics[width=\columnwidth]{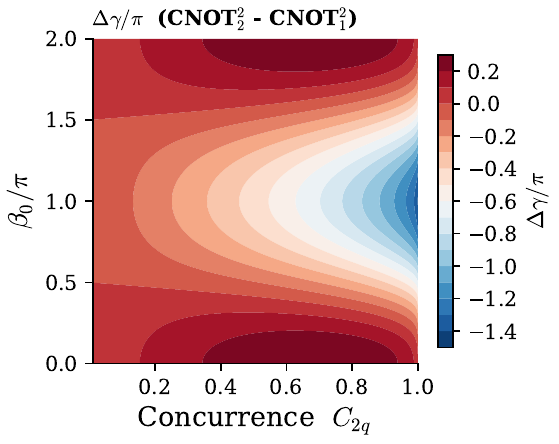}
\caption{Color map of the geometric phase difference $\Delta\gamma/\pi = [\gamma(\mathrm{CNOT}_2^2)-\gamma(\mathrm{CNOT}_1^2)]/\pi$ as a function of concurrence $C_{2q}$ and $\beta_0/\pi$, applied to antisymmetric initial states~\eqref{eq:psi_as}. The sign change and non-monotonic structure reflect the topologically distinct redistribution of geometric phases between the two implementations, confirming that distinction is possible across the full parameter space.}
\label{fig:cnot_phases}
\end{figure}


\noindent where $\op{H}=\frac{1}{\sqrt{2}}(\sigma_x+\sigma_z)$ is the Hadamard operator (Fig.~\ref{fig:cnot_circuit}), implemented by $H_{\op{H}}=E\op{H}$ for time $T=\pi/(2E)$. Applied to antisymmetric states~\eqref{eq:psi_as} over two cycles ($\mathrm{CNOT}^2=I$):
\begin{align}
\gamma({\scriptstyle\mathrm{CNOT}}_1^2) &= \frac{\pi}{2}(\cos\alpha_0 - 1) \nonumber\\
\gamma({\scriptstyle\mathrm{CNOT}}_2^2) &= (1+\sqrt{2})\,\gamma({\scriptstyle\mathrm{CNOT}}_1^2)+\frac{\pi}{\sqrt{2}}\sin\alpha_0\cos\beta_0
\label{eq:CNOT_phases}
\end{align}
The first phase decreases monotonically with concurrence; the second is non-monotonic: it can increase or decrease depending on $\beta_0$. The non-monotonic redistribution of the geometric phase difference $\gamma({\scriptstyle\mathrm{CNOT}}_2^2)-\gamma({\scriptstyle\mathrm{CNOT}}_1^2)$ across the $(\alpha_0,\beta_0)$ parameter space is pictured in fig. \ref{fig:cnot_phases}. It is a direct signature of the identical topological number of CNOT$_i$, $\nu_U=-1$, owing to the fact that $\nu_{H_{\op{H}}}=0$. This is at odds with the usual view that an algorithm is independent of its implementation~\cite{Haroche2006,Burkard2023,Krantz2019,Fang2023}.

\section{Noise Characterization and Hamiltonian Spectroscopy}
\label{sec:noise}

\begin{figure}[t]
\centering
\includegraphics[width=0.78\columnwidth]{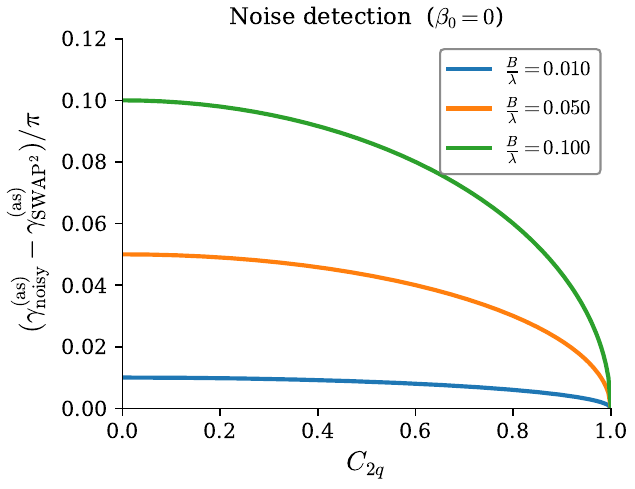}
\caption{Noise detection through the geometric phase. The phase correction $(\gamma_{\mathrm{noisy}}^{(\mathrm{as})} - \gamma_{\mathrm{SWAP}^2}^{(\mathrm{as})})/ \pi$ from~\eqref{eq:gamma3} is plotted as a function of concurrence $C_{2q}=\sin\alpha_0$ for several noise levels $B/\lambda$ (at $\beta_0=0$). %The $B/\lambda=0$ baseline confirms the entanglement-measurement protocol; 
As the noise $B/\lambda$ increases the curve shifts upward linearly, enabling direct quantification of the field from the phase deviation.}
\label{fig:noise}
\end{figure}

The path-sensitivity of the geometric phase also makes it a noise diagnostic. If the Heisenberg gate~\eqref{eq:H2} is perturbed by a small parasitic field ($B/\lambda\ll 1$):
\begin{equation}
H_2 = B\sigma_z^{(1)} + \lambda\,\vec{\sigma}^{(1)}\cdot\vec{\sigma}^{(2)}\,,
\label{eq:H3}
\end{equation}
the geometric phase \eqref{eq:gamma_H2} acquires a linear correction (Fig.~\ref{fig:noise}):
\begin{equation}
\gamma_{\mathrm{noisy}}^{(\mathrm{as})} \approx \gamma_{\op{SWAP}^2}^{(\mathrm{as})}+ \pi\frac{B}{\lambda}\cos\alpha_0\,,
\label{eq:gamma3}
\end{equation}
arises from $H_1$ acting on~\eqref{eq:psi_as}~\cite{Sjoqvist2000}. Scanning $\alpha_0$ or cycle count isolates $B/\lambda$. This notion is not limited to Hamiltonians of the type~\eqref{eq:H3}: although in most cases the noisy path will not be cyclic, the non-cyclic geometric phase converges for cyclic paths~\cite{Sjoqvist2000mixed}, meaning that for small disruptions the noise can be analyzed and quantified in a way similar to~\eqref{eq:gamma3}.

From another point of view, measuring \eqref{eq:gamma3} for an unknown state of the form \eqref{eq:psi_as} using a controlled magnetic field gives access to both non-local variables, the concurrence $C_{2q}(t=0)$ and $\beta_0$, without full state tomography.

\section{Topology as a Necessary Condition for Entanglement}
\label{sec:topology}

The sum rule provides a direct proof that nontrivial topology is necessary for entanglement. Any single-qubit or local Hamiltonian has $n_+=n_-$, hence $\nu_H\equiv0$ and $\nu_U\equiv0$: the geometric phases over any complete basis sum to zero, no redistribution is possible, and no entanglement can be generated (see Appendix for detail). An entangling Hamiltonian must break this eigenvalue balance; the Heisenberg interaction~\eqref{eq:H2} has $\nu_H\neq 0$ for generic couplings. The geometric phases measured throughout this work are thus direct manifestations of this topological imbalance. Examples for this redistribution of geometric phase for different gate implementations are shown in table \ref{table:examples}.

\begin{table}
\begin{tabular}{|c|c|c|c|c|c|c|}
\hline 
Gate & $\gamma_{\left|00\right\rangle }$ & $\gamma_{\left|01\right\rangle }$ & $\gamma_{\left|10\right\rangle }$ & $\gamma_{\left|11\right\rangle }$ & $\sum\gamma$ & $\nu_{U}$\tabularnewline
\hline 
\hline 
SWAP$_{1}$ & $\pi$ & $0$ & $0$ & $\pi$ & $2\pi$ & $+1$\tabularnewline
\hline 
SWAP$_{2}$ & $2\pi$ & $0$ & $0$ & $0$ & $2\pi$ & $+1$\tabularnewline
\hline 
CNOT$_{1}$ & $0$ & $0$ & $-\pi$ & $-\pi$ & $-2\pi$ & $-1$\tabularnewline
\hline 
CNOT$_{2}$ & $2\sqrt{2}\pi$ & $0$ & $-\left(1+\sqrt{2}\right)\pi$ & $-\left(1+\sqrt{2}\right)\pi$ & $-2\pi$ & $-1$\tabularnewline
\hline 
\end{tabular}

\caption{Redistribution of geometric phase $\gamma$ for the computational basis for the SWAP and CNOT implementations presented in the main text.} \label{table:examples}

\end{table}

\section{Discussion}
\label{sec:discussion}
The geometric phase has been extended to mixed states~\cite{Uhlmann1986,Samuel_Bhandari_1988,Sjoqvist2000mixed} and experimentally measured ~\cite{Du2003,Ericsson2005,Ghosh2006,2007nuclear_Berry,Viyuela2018}, so the framework extends naturally to realistic noisy settings. 

The predictions are directly testable on current hardware. On transmon-based superconducting processors~\cite{Krantz2019,Rigetti2010,Malekakhlagh2020}, $T_2$ coherence times exceeding $100\,\mu$s allow hundreds of gate cycles, more than sufficient to resolve the phase differences in~\eqref{eq:SWAP_phases} and~\eqref{eq:CNOT_phases}. In semiconductor spin-qubit platforms~\cite{Loss1998,Burkard2023}, the noise protocol~\eqref{eq:gamma3} directly targets parasitic nuclear-spin field gradients. A non-cyclic counterpart of the entanglement-dependent geometric phase has already been observed in a photonic setting~\cite{Loredo2014}, suggesting the full cyclic protocol is within experimental reach. 

Beyond the applications demonstrated above, our results open several directions. The topological classification of gate Hamiltonians by $\nu_H$ suggests a pre-calibration diagnostic for error mitigation: geometric phase measurements can identify, among equivalent implementations, which topological class is most robust to the dominant noise channel. The sum rule $\nu_U=(1/2\pi)\sum_n\gamma_n$ provides a device-level certification of entanglement capability without state tomography. Finally, decomposing the geometric phase into local and non-local contributions 
%in $(\alpha_0,\beta_0)$ 
constitutes a form of Hamiltonian spectroscopy extendable to multi-qubit processors.

\section{Conclusion}
\label{sec:conclusion}

To summarize, we have established that the topological sum rule $\nu_U=\frac{1}{2\pi}\sum_n\gamma_n = m\nu_H$ governs the distribution of geometric phases across initial states for any gate implementation. Implementations from different topological classes must redistribute phases differently, making their distinction a topological necessity rather than a contingent observation. This framework unifies gate distinction, noise characterization, and entanglement measurement under a single topological principle and explains why nontrivial topology is a prerequisite for quantum entanglement.

\begin{acknowledgments}
This research was funded by the Israel Science Foundation Grant No.~772/21 and the Pazy Foundation. N.O. acknowledges the support received through the VATAT Scholarship Program for PhD Students in Quantum Science and Technology.
\end{acknowledgments}

%%%%%%%%%%%%%%%%%%%%%%%%%%%%%%%%%%%%%%%%%%%%%%%%%%%%%
%%%%%%%%%%%%%%%%%%%%%%%%%%%%%%%%

\appendix

\section{Appendix: Detailed proof of Eq.(1) \label{app:proof}}
\begin{figure*}[t]
  \centering
  \includegraphics[width=0.95\linewidth]{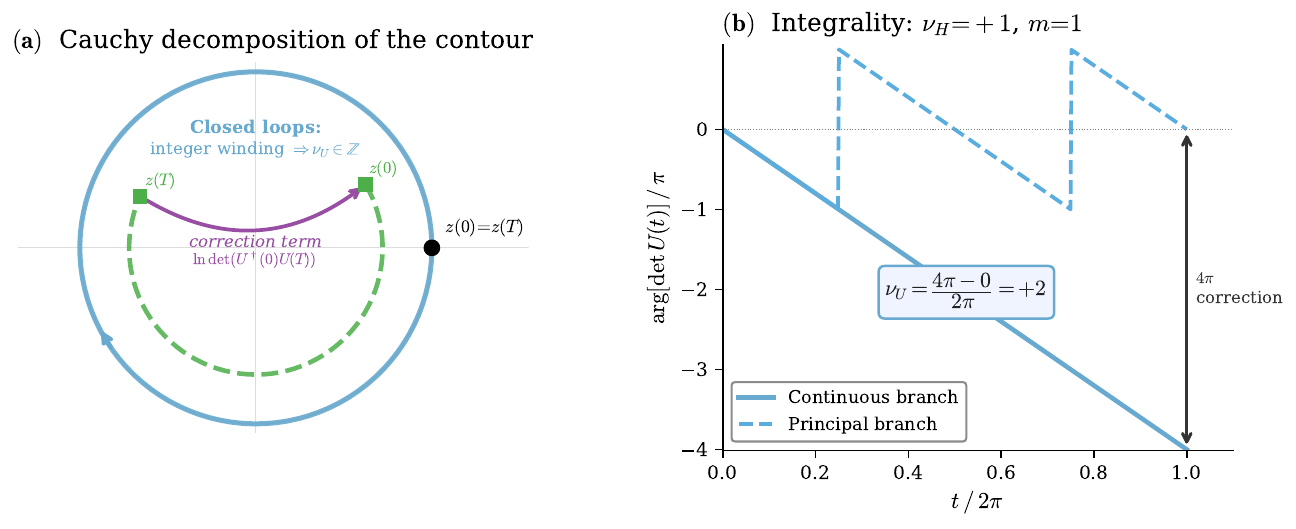}
  \caption{%
    \textbf{(a)}~Cauchy decomposition of the contour: the argument transforming the integral part of~\eqref{eq:nuU} into a Cauchy-type integral.
    \textbf{(b)}~Integrality of $\nu_U$ for $\nu_H=1$ ($n_+-n_-=2$,
    $m=2$): the continuous branch of $\arg(\det U)$ (solid) decreases by
    $4\pi$ over $[0,2\pi]$; the principal branch (dashed) is wrapped
    into $(-\pi,\pi]$. The correction term in Eq.~\eqref{eq:nuU} removes
    the endpoint mismatch, leaving exactly the winding number $\nu_U =m\nu_H = 2$.
  }
  \label{fig:S1}
\end{figure*}
In what follows we show that $\nu_U$~\eqref{eq:nuU} is a gauge-invariant integer (1,\,2), its relation~\eqref{eq:nuU_gamma} with the geometric phase (3), and its relation~\eqref{eq:nuU_nuH} with the Hamiltonian topological number (4). The result $\nu_U=0$ for all local Hamiltonians is explicitly shown (5). We finally explain the method of calculation of $\nu_U$ in a physical scenario, and provide examples (6).

We will utilize the following identity:
\begin{equation}
    \frac{d}{dt}\ln(\det U)=\frac{d}{dt}\text{Tr}\left(\ln U\right)=\text{Tr}\left(U^{\dagger}\frac{d}{dt}U\right) 
\end{equation}
where the last part comes from $U^{-1}=U^{\dagger}$.

\begin{enumerate}
\item Gauge invariance for a small gauge\\
Under $U(t)\to e^{i\rho(t)}U(t)$
the integral part becomes:
\begin{equation}
    \begin{split}
\intop_{0}^{T}\text{Tr}\left(U^{\dagger}\frac{d}{dt}U+\dot{\rho}I\right)dt=&\intop_{0}^{T}\text{Tr}\left(U^{\dagger}\frac{d}{dt}U\right)dt\\ &+\left(\rho\left(T\right)-\rho\left(0\right)\right)\text{Tr}I
\end{split}
\end{equation}
The additional term $\ln\det\!\left(U^{\dagger}(0)\,U(T)\right)=\text{Tr}\left(\ln\left(U^{\dagger}(0)\,U(T)\right)\right)$ becomes:
\begin{equation}
    \begin{split}
 & \rightarrow\text{Tr}\left(\ln\left(e^{i\left(\rho\left(T\right)-\rho\left(0\right)\right)}U^{\dagger}(0)\,U(T)\right)\right)\\
 & =\text{Tr}\left(\ln\left(U^{\dagger}(0)\,U(T)\right)+I\ln e^{i\left(\rho\left(T\right)-\rho\left(0\right)\right)}\right)\\
 & =\left(\ln\left(U^{\dagger}(0)\,U(T)\right)\right)+\left(\rho\left(T\right)-\rho\left(0\right)\right)\text{Tr}I
\end{split}
\end{equation}
Both terms change by the same amount, meaning $\nu_{U}$ is left unchanged. Note that if the natural logarithm is taken in the principle branch, this proof is applicable only if the gauge is small enough, i.e. $\left(\rho\left(T\right)-\rho\left(0\right)\right)\text{Tr}I<2\pi$. See more in (6).

\item Integrality (see Fig.~\ref{fig:S1})\\
The combination in Eq.~\eqref{eq:nuU} equals the winding number of the closed curve $t\mapsto\det U(t)$ around the origin in $\mathbb{C}$. The first term accumulates the total continuous argument of $\det U(t)$; the second subtracts the principal-branch value at the endpoints. Their difference counts how many times $\det U(t)$ encircles the origin and is therefore always an integer.

A different way to look at this proof is by Cauchy's Theorem. Consider the integral part:
\begin{equation}
 \int_{0}^{T}\frac{d}{dt}\ln(\det U)\,dt=\int_{0}^{T}\frac{1}{\det U}\frac{d\left(\det U\right)}{dt}\,dt   
\end{equation}

$\det U(t)$ is a complex function, describing a contour $\Gamma$
in the complex plane of matrix determinants. Exchanging variables
to $z$ we get:
\begin{equation}
    \int_{0}^{T}\frac{1}{\det U}\frac{d\left(\det U\right)}{dt}\,dt=\int_{\Gamma}\frac{1}{z}dz
\end{equation}
The contour is of radius 1, as $|\det U(t)|=1$. For any closed curve
$\int_{\Gamma}\frac{1}{z}dz=0$ as long as $\Gamma$ does not complete
a circle, as $1/z$ is holomorphic on $\mathbb{C}\setminus\{0\}$, so by
Cauchy's theorem, the integral over any closed contour not encircling the origin
vanishes. The non-zero part of the integral can be divided into
two: full circles around $(0,0)$, beginning and ending at $z(0)=\det U(0)$,
and an open contour beginning at $z(0)$ and ending in $z(T)$. The
open contour integral $\int_{z(0)}^{z(T)}\frac{dz}{z}$ exactly cancels
with the second term in $\nu_{U}$. $\nu_{U}$ thus becomes:
\begin{equation}
    \nu_{U}=\frac{i}{2\pi}n\oint\frac{dz}{z}=-n
\end{equation}
where the known Cauchy formula was used, and $n$ is the number of
times $\det U$ winds around $z=0$ clockwise, minus the number of
times it winds anticlockwise, i.e. an integer. This is demonstrated in panel a in Fig.~\ref{fig:S1}. Note that the proof
is independent of $T$, i.e. regardless of periodicity or lack thereof in $U$. 

\item $\nu_{U}$ is the sum of geometric phases.\\
Using the same identity as (1), 
\begin{equation}
    \nu_{U}=\frac{i}{2\pi}\left[\intop_{0}^{T}\text{Tr}\left(U^{\dagger}\frac{d}{dt}U\right)dt-\text{Tr}\left(\ln\left[U^{\dagger}\left(0\right)U\left(T\right)\right]\right)\right]
\end{equation}
We choose to expand the trace in the eigenbasis of $U^{\dagger}\left(0\right)U\left(T\right)$. Since this operator is unitary, its eigenvalues lie on the unit circle: $U^{\dagger}\left(0\right)U\left(T\right)\left|\psi_{n}\right\rangle =e^{i\theta_{n}}\left|\psi_{n}\right\rangle$. This choice is essential for the second term, where passing $\ln$ through the matrix element requires the basis states to be eigenstates. The first term expands as:
\begin{equation}
    \begin{split}
        i\intop_{0}^{T}\text{Tr}\left(U^{\dagger}\frac{d}{dt}U\right)dt= & i\sum_{n}\intop_{0}^{T}\left\langle \psi_{n}\left|U^{\dagger}\frac{d}{dt}U\right|\psi_{n}\right\rangle dt\\
  =&\sum_{n}i\intop_{0}^{T}\left\langle \psi_{n}\left(t\right)\left|\frac{d}{dt}\right|\psi_{n}\left(t\right)\right\rangle dt
    \end{split}
\end{equation}

For the second term, since $\left|\psi_{n}\right\rangle$ are eigenstates of $U^{\dagger}\left(0\right)U\left(T\right)$, we have $\left\langle \psi_{n}\left(0\right)|\psi_{n}\left(T\right)\right\rangle =e^{i\theta_{n}}$ and therefore:
\begin{equation}
    \begin{split}
i\text{Tr}\left(\ln\left[U^{\dagger}\left(0\right)U\left(T\right)\right]\right) & =i\sum_{n}i\theta_{n}\\
 & =i\sum_{n}\ln\left\langle \psi_{n}\left(0\right)|\psi_{n}\left(T\right)\right\rangle \\
 & =-\sum_{n}\arg\left\langle \psi_{n}\left(0\right)|\psi_{n}\left(T\right)\right\rangle    \end{split}
\end{equation}
Combining the two expressions yields:
\begin{equation}
    \begin{split}
\nu_{U}=&\sum_{n}\frac{i}{2\pi}\intop_{0}^{T}\left\langle \psi_{n}\left(t\right)\left|\frac{d}{dt}\right|\psi_{n}\left(t\right)\right\rangle dt\\&+\frac{1}{2\pi}\arg\left\langle \psi_{n}\left(0\right)|\psi_{n}\left(T\right)\right\rangle \\=&\frac{1}{2\pi}\sum_{n}\gamma_{n} \label{eq:nu=sumgamma}
\end{split}
\end{equation}
which is the definition of the geometric phase. Although the derivation uses a specific basis, the final expression $\nu_U = \frac{1}{2\pi}\sum_n \gamma_n$ is basis-independent, since $\nu_U$ itself is defined through the determinant and trace, both of which are basis-independent quantities.

A useful special case arises when $T$ is a shared period of the evolution for all states, i.e.\ $U^{\dagger}\left(0\right)U\left(T\right)=\mathbf{1}$. Then every basis is an eigenbasis, the second term vanishes trivially, and the above derivation holds for any choice of $\left\{\left|\psi_n\right\rangle\right\}$. We have shown in (2) that $\nu_{U}$ is invariant to small perturbations of the energies, meaning such a shared period can always be found.

\item Evaluation on the flattened representative \\
The proof in (2) shows that $\nu_{U}$ is invariant to continuous deformations of $H$ that preserve the signs of eigenvalues, as these will not change the number of times $\det U$
winds around $z=0$. This is a hallmark property of topological invariants, as it means $\nu_U$ is the same for all Hamiltonians in a given topological class. For the flattened representative, $\det U(t) = e^{-2i\nu_H t}$ winds $m\nu_H$ times over the time period $[0,\pi m]$, and $\det U(\pi m) = \mathbf{1}$ (although $U(\pi m)\neq\mathbf{1}$) so the correction term vanishes, giving $\nu_U = m\nu_H$.

\item $\nu_U=0$ for local magnetic fields\\
A local magnetic field $B\hat{u}$ can be described by the Hamiltonian $H_1 = B\hat{u}\cdot\vec{\sigma}^{(1)}$. The resulting phase over a single period $T_1=2\pi/B$ is
\begin{equation}
\gamma_{2q}^{(1)} = 2\pi\sqrt{1-C_{2q}^2}\;\hat{u}\cdot\hat{n}_1(t=0)
\label{eq:gamma_B}
\end{equation}
which is nothing but the single-qubit phase modified by the constant concurrence of the state. It is clear that for every state with Bloch vector $\hat{n}_1$, the orthogonal state $-\hat{n}_1$ contributes a geometric phase of opposite sign, and the sum vanishes.

\item Explicit calculation of $\nu_U$ by~\eqref{eq:nuU} and~\eqref{eq:nu=sumgamma}.\\
The geometric phase $\gamma$ is known to be gauge invariant for small gauge changes, whereas for a so-called "large gauge", $\gamma$ is only invariant modulo $2\pi$ \cite{Mukunda_Simon_1993,Vanderbilt_book_2018}. Consider a gauge transformation $H\rightarrow H+\mu I$:
\begin{equation}
    \begin{split}
         \intop_{0}^{T}i\left\langle \psi\left|\frac{d}{dt}\right|\psi\right\rangle dt &\rightarrow \intop_{0}^{T} i\left\langle \psi\left|\frac{d}{dt}\right|\psi\right\rangle dt +\mu T\\\arg\left\langle \psi\left(0\right)|\psi\left(T\right)\right\rangle &\rightarrow\arg (e^{-i\mu T}\left\langle \psi\left(0\right)|\psi\left(T\right)\right\rangle). \label{eq:gauge_berry}
    \end{split}
\end{equation}
If $\mu T>2\pi$ then as the argument function is resolved in the principle branch, i.e. $\in\{0,2\pi\}$, it cannot correct the change in the first term.

The same non-invariance applies to $\nu_U$ through $\gamma_n$ in~\eqref{eq:nu=sumgamma} or by the $\ln\det\!\left(U^\dagger(0)\,U(T)\right)$ term in~\eqref{eq:nuU}. This conundrum can be solved by looking at the topological number of the Hamiltonian: it is only gauge invariant as long as the imbalance between positive and negative eigenvalues is unchanged. $\nu_H$ was calculated based on a traceless Hamiltonian (equation (2) on the main text), thus all relations presented a valid for any gauge that respects $\nu_H$.

In short, calculation of $\gamma$ and $\nu_U$ can be made as usual if the governing Hamiltonian is traceless or close to traceless. When $T\left|\op{Tr}(H)\right|>2\pi$ where $T$ is the period, then the calculated $\gamma$ should be corrected by adding the term 
\begin{equation}
    \Delta\gamma_{\text{cor}}=-2\pi\op{sgn}(\text{Tr}\left(H\right))\left\lfloor \frac{T\left|\text{Tr}\left(H\right)\right|}{2\pi}\right\rfloor.
\end{equation}
This can be seen by look at~\eqref{eq:gauge_berry}. The addition of the two terms gives an extra $+\mu T-(\mu T \mod 2\pi)$, which is just $2\pi$ times the amount of times $2\pi$ enters in $T\left|\text{Tr}\left(H\right)\right|$, with the appropriate sign.
In accordance with $\gamma$, $\nu_U$ should be corrected by adding $\Delta \nu
_U=S\Delta\gamma_{\text{cor}}$ where $S$ is the size of the matrix $H$ (or $U$).

\end{enumerate}

\bibliography{Bibliography/proposal}

\end{document}

\title{Supplemental Material for ``Topological sum rule for geometric phases of quantum gates''}

\author{Nadav Orion}
\affiliation{Department of Physics, Technion, Israel Institute of Technology, Haifa 3200003, Israel}

\author{Boris Rotstein}
\affiliation{Department of Physics, Technion, Israel Institute of Technology, Haifa 3200003, Israel}

\author{Nirron Miller}
\affiliation{Department of Physics, The University of Texas, Austin, Texas 78712}

\author{Eric Akkermans}
\email{eric@physics.technion.ac.il}
\affiliation{Department of Physics, Technion, Israel Institute of Technology, Haifa 3200003, Israel}

\date{\today}

\maketitle
%%%%%%%%%%%%%%%%%%%%%%%%%%%%%%%%%%%%%%%%%%%%%%%%%%%%%%%%%%%%%
\section{State representation by physical parameters}
In this section, we show that a general two-qubit state, up to a total phase, can be written as 
\begin{equation}
    \left|\psi_{2q}\right>=\cos\frac{\alpha}{2}e^{-i\beta/2}\left|\hat{n}_1,\hat{n}_2\right>+\sin\frac{\alpha}{2}e^{i\beta/2} \left|-\hat{n}_1,-\hat{n}_2\right> 
    \label{eq:general_2q}
\end{equation}
where
\begin{equation}
    \begin{split}
    \left|\hat{n}_i\right>&=\cos\frac{\theta_i}{2}e^{-i\phi_i/2}\left|0\right>+\sin\frac{\theta_i}{2}e^{i\phi_i/2} \left|1\right>\\
    \left|-\hat{n}_i\right>&=-\sin\frac{\theta_i}{2}e^{-i\phi_i/2}\left|0\right>+\cos\frac{\theta_i}{2} e^{i\phi_i/2}\left|1\right> \label{eq:1q_basis}
    \end{split}
\end{equation}
and derive the bounds for $\alpha$ and $\beta$.

We begin with the Schmidt decomposition of a two-qubit state:
\begin{equation}
    \left|\psi\right\rangle =\lambda_{+}\left|\psi_{1}^{+}\right\rangle \otimes\left|\psi_{2}^{+}\right\rangle +\lambda_{-}\left|\psi_{1}^{-}\right\rangle \otimes\left|\psi_{2}^{-}\right\rangle 
\end{equation}
 where 
 \begin{equation}
     \left\langle \psi_{i}^{m}|\psi_{i}^{n}\right\rangle =\delta_{mn}, \label{1q_orthonomral}
 \end{equation}
$\lambda_\pm\in[0,1]$ and $\lambda_{-}=\sqrt{1-\lambda_{+}^2}$. There is a small redundancy here, as the cases of $\lambda_+>\sqrt{2}/2$ are equivalent to changing the name $+\leftrightarrow-$ and $\lambda_+\rightarrow\sqrt{1-\lambda_{+}^2}$.
Thus we choose $\lambda_+\in[0,\sqrt{2}/2]$. We can now denote $\lambda_+=\cos\frac{\alpha}{2}$ where $\alpha\in[0,\pi/2]$.

Next, by \eqref{1q_orthonomral}, $\left|\psi_{i}^{+}\right\rangle$ and $\left|\psi_{i}^{-}\right\rangle$ are a basis of the single-qubit Hilbert space of qubit $i$. Any such basis can canonically be written as the eigenstate of some $\hat{n}_i\cdot\vec{S}_i$:
\begin{equation}
 \begin{split}
    \left|\psi_{i}^{+}\right\rangle &=e^{i\mu_{i}}\left[\cos\frac{\theta_{i}}{2}e^{-i\phi_{i}/2}\left|0\right>+\sin\frac{\theta_{i}}{2}e^{i\phi_{i}/2}\left|1\right>\right]\equiv e^{i\mu_{i}}\left|\hat{n}_{i}\right>\\
    \left|\psi_{i}^{-}\right\rangle &\equiv e^{i\rho_{i}}\left|-\hat{n}_{i}\right>
    \end{split}
\end{equation}
Such that
\begin{equation}
 \left|\psi\right\rangle =e^{i\left(\mu_{1}+\mu_{2}\right)}\cos\frac{\alpha}{2}\left|\hat{n}_{1},\hat{n}_{2}\right\rangle +e^{i\left(\rho_{1}+\rho_{2}\right)}\sin\frac{\alpha}{2}\left|-\hat{n}_{1},-\hat{n}_{2}\right\rangle    
\end{equation}

Denoting $\beta\equiv\rho_{1}+\rho_{2}-\mu_{1}-\mu_{2}$, up to a total phase
\begin{equation}
    \left|\psi\right>=\cos\frac{\alpha}{2}e^{-i\beta/2}\left|\hat{n}_{1},\hat{n}_{2}\right>+\sin\frac{\alpha}{2}e^{i\beta/2}\left|-\hat{n}_{1},-\hat{n}_{2}\right> \label{eq:psi2q}
\end{equation}

%%%%%%%%%%%%%%%%%%%%%%%%%%%%%%%%%%%%%%%%%%%%%%%%%%%%%%%%%%

\section{Physical meaning of $\alpha$ and $\beta$, and rotation representation}
To clarify the functions of $\beta$ and $\alpha$, let us first introduce an effective spin algebra in the Schmidt sphere. Looking at  \eqref{eq:general_2q}, we rename $\left|\Uparrow\right\rangle\equiv\left|\hat{n}_1\right>\otimes\left|\hat{n}_2\right>$ and $\left|\Downarrow\right\rangle\equiv\left|-\hat{n}_1\right>\otimes\left|-\hat{n}_2\right>$ and define:

\begin{equation}
\begin{array}{c}
\tilde{\Sigma}_{x}=\left|\Uparrow\right\rangle \left\langle \Downarrow\right|+\left|\Downarrow\right\rangle \left\langle \Uparrow\right| ,\quad \tilde{\Sigma}_{y}=-i\left|\Uparrow\right\rangle \left\langle \Downarrow\right|+i\left|\Downarrow\right\rangle \left\langle \Uparrow\right|
\\
\\
\tilde{\Sigma}_{z}=\left|\Uparrow\right\rangle \left\langle \Uparrow\right|-\left|\Downarrow\right\rangle \left\langle \Downarrow\right| ,\quad \tilde{I}=\left|\Uparrow\right\rangle \left\langle \Uparrow\right|+\left|\Downarrow\right\rangle \left\langle \Downarrow\right|
\label{tilde}
\end{array}
\end{equation}
\begin{equation}
\left[\tilde{\Sigma}_{j},\tilde{\Sigma}_{k}\right]=2i\varepsilon_{jk\ell}\tilde{\Sigma}_{\ell}. \label{tilde_algebra}
\end{equation}
The generators of this algebra are non-local and their expectation values depend on $\alpha$ and $\beta$ only:
\begin{equation}
    \left\langle \psi_{2q}\left|\tilde{\boldsymbol{\Sigma}}\right|\psi_{2q}\right\rangle  = \left(\begin{array}{c}
\sin\alpha\cos\beta\\
\sin\alpha\sin\beta\\
\cos\alpha
\end{array}\right) \label{eq:effective_qubit_direction}
\end{equation}
Observe the resemblance to single-qubit expectation values. The angles $\alpha$ and $\beta$ create an additional, non-local ``Bloch sphere'' alongside those for each qubit. The state vector's projection on the $xy$ plane directly corresponds to the concurrence $\sin{\alpha}=C_{2q}$, as shown in the main text.
As anticipated, $\beta$ plays a role in non-local measurements.

The $\tilde{\Sigma}$ operators can be related to the Pauli matrices in any chosen basis. For example, taking $\left|\Uparrow\right\rangle\equiv\left|00\right>$ and $\left|\Downarrow\right\rangle\equiv\left|11\right>$ results in:
\begin{equation}
   \tilde{\boldsymbol{\Sigma}}_{\ket{00},\ket{11}} = \left(\begin{array}{c}
\frac{1}{2}\left(\sigma_{x}\otimes\sigma_{x}-\sigma_{y}\otimes\sigma_{y}\right)\\
\frac{1}{2}\left(\sigma_{x}\otimes\sigma_{y}+\sigma_{y}\otimes\sigma_{x}\right)\\
\frac{1}{2}\left(\sigma_{z}\otimes I+ I\otimes\sigma_{z}\right)
\end{array}\right).
\label{tilde_AS}
\end{equation}
For the case of qubits in general directions $\hat{n}_{1,2}$, each Pauli matrix is replaced by its equivalent in the rotated $\hat{n}_{1,2}$ basis \eqref{eq:1q_basis}. For example if $\hat{n}_1\equiv M\hat{z}$, where $M$ is a rotation matrix of $\mathbb{R}^3$, then $\sigma_i \otimes I \rightarrow (M\hat{i})\cdot\vec\sigma\otimes I$ for $i\in\{x,y,z\}$, e.g. $\sigma_z\otimes I \rightarrow \hat{n}_1\cdot\vec\sigma\otimes I$. 

%%%%%%%%%%%%%%%%%%%%%%%%%%%%%%%%%%%%%%%%%%%%%%%%%%%%%%%%%%%%%

\section{Derivation of geometric phase for two qubits and discussion}

In this section we derive the Berry connection leading to the two-qubit geometric phase formula of the main text. Following \cite{Bouchiat1988}, the general state \eqref{eq:general_2q} can be expressed as a product of three rotations of the state $\left|00\right\rangle$:
\begin{equation}
    \begin{array}{rrrr}
    \left|\psi\right\rangle =R_{1}\left(\theta_{1},\varphi_{1}\right)R_{2}\left(\theta_{2},\varphi_{2}\right)R_{3}\left(\alpha,\beta\right)\left|00\right\rangle \\ 
    R_{1}=\exp{(-i\frac{\varphi_{1}}{2}\sigma_{z}^{(1)})}\exp{(-i\frac{\theta_{1}}{2}\sigma_{y}^{(1)})}\\ 
    R_{2}=\exp{(-i\frac{\varphi_{2}}{2}\sigma_{z}^{(2)})} \exp{(-i\frac{\theta_{2}}{2}\sigma_{y}^{(2)})} \\ 
    R_{3}=\exp{(-i\frac{\beta}{2}\tilde{\Sigma}_{z})}\exp{(-i\frac{\alpha}{2}\tilde{\Sigma}_{y})}.
    \end{array}
    \label{eq:rotation_matrices}
\end{equation}
where $\left|\Uparrow\right\rangle \equiv\left|00\right\rangle $
and $\left|\Downarrow\right\rangle \equiv\left|11\right\rangle $ for the $\tilde{\Sigma}_i$ operators. Put into words,
$R_{1}$ ($R_{2}$) determines the direction of qubit 1 (2), and $R_3$ determines the direction in the Schmidt sphere, as explained in the previous section.

The Berry connection $\mathcal{A}$ takes the form:
\begin{equation}
\begin{split}
\left\langle \psi\left|d\right|\psi\right\rangle  & =\left\langle 00\left|\left(R\right)^{-1}dR\right|00\right\rangle \\
 & =\left\langle 00\left|R_{3}^{-1}R_{1}^{-1}dR_{1}R_{3}\right|00\right\rangle \\
 &+\left\langle 00\left|R_{3}^{-1}R_{2}^{-1}dR_{2}R_{3}\right|00\right\rangle
  \\
 & +\left\langle 00\left|R_{3}^{-1}dR_{3}\right|00\right\rangle,
\end{split}
 \label{eq:terms}
\end{equation}
where we used $\left[R_{1},R_{2}\right]=0$, and $\left[R_{1},\dot{R}_{2}\right]=\left[\dot{R}_{1},R_{2}\right]=0$
as they act on orthogonal spaces. This expression can be further simplified using rotation-matrix identities:
\begin{align}
R^{-1}\left(\vec{n},\alpha\right)\cdot dR=R\left(\vec{n},d\alpha\right)-I \label{id1}\\
R^{\prime-1}R\left(\vec{n},\alpha\right)R^{\prime}=R\left(R^{\prime-1}\vec{n},\alpha\right)\label{id2}
\end{align}
which will be proven in the next section. Using identities \eqref{id1} and \eqref{id2}:
\begin{equation}
\begin{split}
R_{1}^{-1}dR_{1} =& R_{1}^{-1}\left(y,\theta_{1}\right)R_{1}^{-1}\left(z,\varphi_{1}\right)\left(dR_{1}\left(z,\varphi_{1}\right)\right)R_{1}\left(y,\theta_{1}\right)\\
 +& R_{1}^{-1}\left(y,\theta_{1}\right)R_{1}^{-1}\left(z,\varphi_{1}\right)R_{1}\left(z,\varphi_{1}\right)\left(dR_{1}\left(y,\theta_{1}\right)\right)\\
 =& R_{1}^{-1}\left(y,\theta_{1}\right)R_{1}\left(z,d\varphi_{1}\right)R_{1}\left(y,\theta_{1}\right) \\
 + & R_{1}\left(y,d\theta_{1}\right)-2I\\
 =& R_{1}\left(R_{1}^{-1}\left(y,\theta_{1}\right)\hat{z},d\varphi_{1}\right)+R_{1}\left(y,d\theta_{1}\right)-2I
\end{split}
\end{equation}

Note that the rotation of the axis $\hat{z}$ is a rotation in $\mathbb{R}^3$ and not a rotation of a spinor. 
\[
R_{1}^{-1}\left(y,\theta_{1}\right)\hat{z}=R_{1}\left(y,-\theta_{1}\right)\hat{z}=\hat{x}\sin\theta_{1}+\hat{z}\cos\theta_{1}
\]
\begin{equation}
\begin{split}
R_{1}^{-1}dR_{1} =& R_{1}\left(\hat{x}\sin\theta_{1}+\hat{z}\cos\theta_{1},d\varphi_{1}\right)+R_{1}\left(y,d\theta_{1}\right)-2I\\
 =& -\frac{i}{2}\left(\sigma_{x}^{(1)}\sin\theta_{1}+\sigma_{z}^{(1)}\cos\theta_{1}\right)d\varphi_{1}-\frac{i}{2}\sigma_{y}^{(1)}d\theta_{1} \label{eq:R1R1}
\end{split}
\end{equation}

Using the definition of $R_3$ from equation \eqref{eq:rotation_matrices}:
\begin{equation}
 \begin{split}
  R_{3}\left|00\right\rangle & =e^{-i\frac{\beta}{2}\tilde{\Sigma}_{z}}e^{-i\frac{\alpha}{2}\tilde{\Sigma}_{y}}\left|00\right\rangle \\
  & =e^{-i\beta/2}\cos\frac{\alpha}{2}\left|00\right\rangle +e^{i\beta/2}\sin\frac{\alpha}{2}\left|11\right\rangle , 
\end{split}   
\end{equation}
and equation \eqref{eq:R1R1} we expand the first term in equation \eqref{eq:terms}, such that:
\begin{equation}
\begin{split}
\left\langle 00\left|R_{3}^{-1}R_{1}^{-1}dR_{1}R_{3}\right|00\right\rangle =&-\frac{i}{2}\cos^{2}\frac{\alpha}{2}d\varphi_{1}\left\langle 0\left|\sigma_{x}\sin\theta_{1}\right|0\right\rangle \\&-\frac{i}{2}\cos^{2}\frac{\alpha}{2}d\varphi_{1}\left\langle 0\left|\sigma_{z}\cos\theta_{1}\right|0\right\rangle \\&-\frac{i}{2}\cos^{2}\frac{\alpha}{2}d\theta_{1}\left\langle 0\left|\sigma_{y}\right|0\right\rangle \\&-\frac{i}{2}\sin^{2}\frac{\alpha}{2}d\varphi_{1}\left\langle 1\left|\sigma_{x}\sin\theta_{1}\right|1\right\rangle \\&-\frac{i}{2}\sin^{2}\frac{\alpha}{2}d\varphi_{1}\left\langle 1\left|\sigma_{z}\cos\theta_{1}\right|1\right\rangle \\&-\frac{i}{2}\sin^{2}\frac{\alpha}{2}d\theta_{1}\left\langle 1\left|\sigma_{y}\right|1\right\rangle \\&=-\frac{i}{2}\cos\alpha\cos\theta_{1}d\varphi_{1}
\end{split}
 \label{term1}
\end{equation}

Via similar arguments,
\begin{equation}
\left\langle 00\left|R_{3}^{-1}R_{2}^{-1}dR_{2}R_{3}\right|00\right\rangle =-\frac{i}{2}\cos\alpha\cos\theta_{2}d\varphi_{2} \label{term2}
\end{equation}

Using the rotation identities for $R_3$ gives
\begin{equation}
R_{3}^{-1}dR_{3} = -\frac{i}{2}\left(\tilde{\Sigma}_{x}\sin\alpha+\tilde{\Sigma}_{z}\cos\alpha\right)d\beta-\frac{i}{2}\tilde{\Sigma}_{y}d\alpha
\end{equation}
such that the last term in equation \eqref{eq:terms} is:
\begin{equation}
\begin{split}
\left\langle 00\left|R_{3}^{-1}dR_{3}\right|00\right\rangle =&-\frac{i}{2}\left\langle 00\right|\left(\tilde{\Sigma}_{x}\sin\alpha+\tilde{\Sigma}_{z}\cos\alpha\right)d\beta\left|00\right\rangle \\&-\left\langle 00\right|\frac{i}{2}\tilde{\Sigma}_{y}d\alpha\left|00\right\rangle \\=&-\frac{i}{2}\cos\alpha d\beta
 \label{term3}
\end{split}
\end{equation}

Gathering all the terms gives:
\begin{equation}
\left\langle \psi\left|d\right|\psi\right\rangle =-\frac{i}{2}\cos\alpha\left(d\beta+\cos\theta_{1}d\varphi_{1}+\cos\theta_{2}d\varphi_{2}\right) \label{eq:BerryA}.
\end{equation}

To obtain the geometric phase we now integrate $\left\langle \psi|d\psi\right\rangle$ on a chosen path $\Gamma$ in the Hilbert space. We have chosen a time-independent parametrization for each point in the Hilbert space, given by the representative $\ket{\psi}$ \eqref{eq:rotation_matrices}. This constitutes a time-independent gauge. In other words, when a path is closed in the parameter space $\alpha,\beta,\theta_{1,2},\phi_{1,2}$ then $\langle\psi(\text{start})|\psi(\text{end})\rangle=1$  by design, i.e.  $\arg\langle\psi(\text{start})|\psi(\text{start})\rangle=0$. Hence the geometric phase is given by:
\begin{equation}
    \gamma_{2q} = \frac{1}{2}\int_\Gamma\,\sqrt{1-C_{2q}^2}\left(d{\beta}+d{\phi}_1\cos\theta_1 + d{\phi}_2\cos\theta_2\right)
\label{eq:gamma2q_path}
\end{equation}
where we used $\cos\alpha=\sqrt{1-C_{2q}^2}$.
We now parametrize the path $\Gamma$ by a single time parameter $t$.The additional term $\arg\langle\psi(\text{start})|\psi(\text{start})\rangle=\arg\langle\psi(0)|\psi(T)\rangle$ vanishes by the same argument, giving:
\begin{equation}
    \gamma_{2q} = \frac{1}{2}\int_0^T dt\,\sqrt{1-C_{2q}^2}\left(\dot{\beta}+\dot{\phi}_1\cos\theta_1 + \dot{\phi}_2\cos\theta_2\right). \label{eq:gamma2q}
\end{equation}
This is the two-qubit geometric phase formula cited in Eq.~(9) of the main text.

The single-qubit geometric phase $\int dt\dot{\phi}\cos\theta$~\cite{Aharonov1987} appears twice in the two-qubit expression~\eqref{eq:gamma2q}, contributing once for each qubit. This result arises directly from the physical meaning of $\hat{n}_i$ as the observation value of the local spin operators, i.e.
\begin{equation}
\langle\vec{\sigma}^{(i)}\rangle = \hat{n}_i\cos\alpha
\label{eq:sigmaexp}
\end{equation}
% \begin{equation}
% \langle\vec{\sigma}^{(1)}\rangle = \hat{n}_1\cos\alpha\,,\qquad \langle\vec{\sigma}^{(2)}\rangle = \hat{n}_2\cos\alpha
% \label{eq:sigmaexp}
% \end{equation}
where $\vec{\sigma}^{(i)}$ denotes $\vec{\sigma}$ acting locally on qubit $i$. Measuring the qubit $i$ along the $+\hat{n}_i$ ($-\hat{n}_i$) direction, yields $+$ (resp. $-$) with probability $\cos^2(\alpha/2)$ (resp. $\sin^2(\alpha/2)$).

At first glance,~\eqref{eq:sigmaexp} might suggest that $\alpha$ (and therefore $C_{2q}$) is a local quantity, but its accessibility via local measurements arises from the restriction to pure states. In mixed states both entanglement and noise (specifically depolarization) contribute towards the size of $\langle\vec{\sigma}^{(i)}\rangle$. Local measurements cannot differentiate between the two. The physical significance of the phase $\beta$ is less intuitively apparent, since it is generated locally and does not depend on entanglement. At the same time, as implied by~\eqref{eq:sigmaexp}, $\beta$ cannot be inferred using only local operations (see SM). The structure generated by $(\alpha,\beta)$ is called the Schmidt sphere~\cite{Sjoqvist2000}, although it is not strictly a sphere.

There are two key differences between~\eqref{eq:gamma2q} and the single-qubit result.
%Alternative: The smeblance between ~\eqref{eq:gamma2q} and the single-qubit result breaks at two key points. 
First, the single-qubit phase is modified by the concurrence, as reflected in the $\alpha$-dependence of the Bloch vector in~\eqref{eq:sigmaexp}. In the absence of entanglement, $C_{2q}\equiv 0$, and~\eqref{eq:gamma2q} reduces to two independent single-qubit contributions, $\gamma_{2q}=\sum_i\gamma_{1q_i}$~\footnote{The $\dot{\beta}$ term in~\eqref{eq:gamma2q} vanishes, as in that case~\eqref{eq:psi2q} is a separable state, and $\beta$ is a global phase.}. Second, and more substantially,~\eqref{eq:gamma2q} contains an extra term $\sqrt{1-C_{2q}^2}\,\dot{\beta}$ with no single-qubit analog, arising from inter-qubit interactions and thus non-local in nature.

The role of this new term can be clarified by an example. Consider the Heisenberg-type interaction between the two qubits, mediated by the exchange term:
\begin{equation}
H_2 = \lambda\,\vec{\sigma}^{(1)}\cdot\vec{\sigma}^{(2)} = \lambda(\sigma_x\otimes\sigma_x + \sigma_y\otimes\sigma_y + \sigma_z\otimes\sigma_z)
\label{eq:H2}
\end{equation}
while considering antisymmetric initial states,
\begin{equation}
|\psi_{\mathrm{as}}(t\!=\!0)\rangle = \cos\frac{\alpha_0}{2}\,e^{-i\beta_0/2}|10\rangle + \sin\frac{\alpha_0}{2}\,e^{i\beta_0/2}|01\rangle\,.
\label{eq:psi_as}
\end{equation}
It has been shown~\cite{Sjoqvist2000} that $H_2$ acting on the subspace defined by~\eqref{eq:psi_as} is equivalent to rotation around the $\hat{x}$ direction of the $(\alpha,\beta)$ Schmidt sphere, similar to a magnetic field in the $\hat{x}$ direction for a qubit. The time evolution is periodic with $T_2=\pi/\lambda$, and the resulting geometric phase for a single period is:
\begin{equation}
\gamma_{2q}^{(1)} = 2\pi\sin\alpha_0\cos\beta_0 = 2\pi\,C_{2q}(t\!=\!0)\cos\beta_0\,.
\label{eq:gamma_H2}
\end{equation}
This result isreminiscent of a single qubit in a magnetic field, where $\sin\alpha_0\cos\beta_0$ is the $\hat{x}$ component of the effective qubit on the Schmidt sphere. Note that no magnetic field is required to produce the geometric phase, and neither qubit changes its orientation on the Bloch sphere. In other words, $\gamma_{2q}^{(1)}$ arises solely from the $\dot{\beta}$ term in~\eqref{eq:gamma2q}. This is in concordance with $H_2$ interpreted as a rotation in the Schmidt sphere \cite{Sjoqvist2000}. 
%%%%%%%%%%%%%%%%%%%%%%%%%%%%%%%%%%%%%%%%%%%%%%%%%%%%%%%%%%%%%

\section{Proof of rotation identities}
In this section we prove identities \eqref{id1} and \eqref{id2}. Proving identity \eqref{id1}:
\begin{equation}
\begin{split}
R^{-1}dR=&d\alpha e^{-\frac{1}{2}i\vec{n}\cdot\vec{\sigma}\alpha}\frac{d}{d\alpha}e^{\frac{1}{2}i\vec{n}\cdot\vec{\sigma}\alpha}\\=&d\alpha e^{-\frac{1}{2}i\vec{n}\cdot\vec{\sigma}\alpha}\frac{d}{d\alpha}\left(I\cos\frac{\alpha}{2}+i\vec{n}\cdot\vec{\sigma}\sin\frac{\alpha}{2}\right)\\=&\frac{1}{2}d\alpha\left(I\cos\frac{\alpha}{2}-i\vec{n}\cdot\vec{\sigma}\sin\frac{\alpha}{2}\right)\cdot\\&\cdot\left(-I\sin\frac{\alpha}{2}+i\vec{n}\cdot\vec{\sigma}\cos\frac{\alpha}{2}\right)\\=&\frac{1}{2}i\vec{n}\cdot\vec{\sigma}d\alpha\\=&I\cos\frac{d\alpha}{2}+i\vec{n}\cdot\vec{\sigma}\sin\frac{d\alpha}{2}-I\\=&R\left(\vec{n},d\alpha\right)-I
\end{split}
\end{equation}

We shall only use a specific form of \eqref{id2}:
\begin{equation}
R^{\prime-1}\left(y,\theta\right)R\left(z,\varphi\right)R^{\prime}\left(y,\theta\right)=R_{1}\left(\hat{x}\sin\theta+\hat{z}\cos\theta,\varphi\right) \label{id2.5}
\end{equation} 
and its proof:
\begin{equation}
\begin{split}
R^{\prime-1}RR^{\prime} & =e^{\frac{1}{2}i\sigma_{y}\theta}e^{-\frac{1}{2}i\sigma_{z}\varphi}e^{-\frac{1}{2}i\sigma_{y}\theta}\\
 & =e^{\frac{1}{2}i\sigma_{y}\theta}\left(I\cos\frac{\varphi}{2}-i\sigma_{z}\sin\frac{\varphi}{2}\right)e^{-\frac{1}{2}i\sigma_{y}\theta}\\
 & =I\cos\frac{\varphi}{2}-i\left(\sigma_{z}\cos\theta+\sigma_{x}\sin\theta\right)\sin\frac{\varphi}{2}\\
 & =e^{-\frac{1}{2}i\left(\hat{x}\sin\theta+\hat{z}\cos\theta\right)\cdot\vec{\sigma}\varphi}\\
 & =R\left(\hat{x}\sin\theta+\hat{z}\cos\theta,\varphi\right)
\end{split}
\end{equation}

%%%%%%%%%%%%%%%%%%%%%%%%%%%%%%%%%%%%%%%%%%%%%%%%%%%%%%%%%%%%%

\section{Phase summation in quantum circuits}
In this section the method for computing geometric phase in a quantum circuit is presented.

A quantum gate $G$ is implemented by applying a specific Hamiltonian $H$ for a specific time period $\tau$, i.e.  $G=e^{-iH\tau}$. A series of quantum gates is thus modeled by a piecewise time evolutions, each with a different Hamiltonian $H_n$ and with the initial state equal to the final state of the previous piece: 

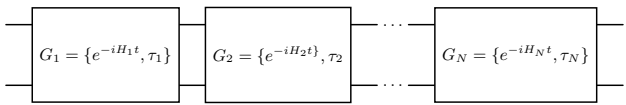
\begin{figure}[ht]
\begin{tikzpicture}
\node[scale=0.7] {
 \begin{quantikz}
&\gate[2]{G_1=\{e^{-iH_1 t},\tau_1\}}&\gate[2]{G_2=\{e^{-iH_2 t\}},\tau_2}&\ \ldots\ &\gate[2]{G_N=\{e^{-iH_N t},\tau_N\}}&\\
&&& \ \ldots\ &&
\end{quantikz}
};
\end{tikzpicture}
	\caption{A two-qubit quantum circuit consisting of $N$ gates, each with its Hamiltonian $H_n$ and gate-time $\tau_n$.}
	\label{sfig:general_circuit}
\end{figure}

The system state at time $t$ is thus
\begin{equation}
\left|\psi\left(t\right)\right\rangle =\begin{cases}
e^{-iH_{1}t}\left|\psi_{0}\right\rangle  & 0<t<\tau_{1}\\
e^{-iH_{2}\left(t-\tau_{1}\right)}G_{1}\left|\psi_{0}\right\rangle  & \tau_{1}<t<\tau_{1}+\tau_{2}\\
\vdots\\
e^{-iH_{N}\left(t-T_{N-1}\right)}\prod_{i=1}^{N-1}G_{i}\left|\psi_{0}\right\rangle  & T_{N-1}<t<T_{N}
\end{cases}
\end{equation}
where $G_n=e^{-iH_{n}\tau_n}$, and $T_{n-1}=\sum_{i=1}^{n-1}\tau_i$ such that the initial state input of gate $n$ is $\left|\psi\left(T_{n-1}\right)\right\rangle =\prod_{i=1}^{n-1}G_{i}\left|\psi_{0}\right\rangle $. 

If the evolution is cyclic up to a phase, i.e. $\left|\psi\left(T_N\right)\right\rangle=e^{i\chi}\left|\psi(t=0)\right\rangle$, the geometric phase gathered by $\left|\psi\left(t\right)\right\rangle$ is given by
\begin{equation}
    \begin{split}
        \gamma&=\chi+i\intop_{0}^{T_{N}}dt\left\langle \psi\left(t\right)\left|\frac{d}{dt}\right|\psi\left(t\right)\right\rangle \\&=\arg\left\langle \psi_{0}|\psi\left(T_{N}\right)\right\rangle +\sum_{n=1}^{N}i\intop_{T_{n-1}}^{T_{n}}dt\left\langle \psi\left(t\right)\left|\frac{d}{dt}\right|\psi\left(t\right)\right\rangle 
    \end{split}
\end{equation}
meaning that the dynamic part is just a piecewise sum of the phases gathered for each Hamiltonian, but the total phase is determined only at the end of the evolution. 

In this section the change between Hamiltonians was assumed to be abrupt, but as shown in Sec.~VI, a slow change will not change the main result.

%%%%%%%%%%%%%%%%%%%%%%%%%%%%%%%%%%%%%%%%%%%%%%%%%%%%%%%%%%%%%

\section{Ramp-up gates}
In the article we model a gate by abrupt Hamiltonians appearing and disappearing suddenly. We will now show that a more physical approach, using a gradual ramp-up, does not change our results. The gate Hamiltonian is modeled as multiplied by a time-dependent function $\lambda\left(t\right)$:
\[
H\left(t\right)=\lambda\left(t\right)H
\]
where $\lambda(t)$ is the dimensionless Hamiltonian strength. In an experiment we expect it to be a variant of a tent function. The time-dependent Schr\"odinger equation is
\[
i\frac{d}{dt}\left|\psi\right\rangle =\lambda\left(t\right)H\left|\psi\right\rangle.
\]
The eigenstates are constant in time, so as long as there is no degeneracy:
\[
\left|\psi\left(t\right)\right\rangle =\sum c_{n}\left(t\right)\left|n\right\rangle 
\]
where $\left|n\right\rangle $ are the eigenstates of $H$.
\[
i\dot{c}_{n}=\lambda\left(t\right)E_{n}c_{n}
\]
\[
c_{n}=\exp\left(iE_{n}\intop_{0}^{t}dt'\lambda\left(t'\right)\right)
\]
if $\lambda\left(t\right)>0\;\forall\, t$,
then this is equivalent to a new time parameter $\tau\left(t\right)=\intop_{0}^{t}dt'\lambda\left(t'\right)$. Thus, the time evolution of a state can be described using a time-independent Hamiltonian with a new time parameter.
The geometric phase does not depend on time, only on the path in the
Hilbert space. Thus, application of a gate abruptly or by ``ramping
up and down'' will not change our results.

%%%%%%%%%%%%%%%%%%%%%%%%%%%%%%%%%%%%%%%%%%%%%%%%%%%%%%%%%%%%%

\bibliography{Bibliography/proposal}